\documentclass[a4paper]{article}
\usepackage{listings}
\usepackage{xcolor}
\usepackage{multirow}
\usepackage{jheppub}  
\usepackage{dcolumn}
\usepackage{placeins}
\usepackage[english]{babel}
\usepackage[utf8x]{inputenc}
\usepackage[T1]{fontenc}
\usepackage{palatino}
\usepackage[switch*]{lineno}
\usepackage{amsmath}
\usepackage{afterpage}
\usepackage{graphicx, subcaption}
\usepackage[colorinlistoftodos]{todonotes}
\usepackage[colorlinks=true]{hyperref}
\usepackage{makeidx}
\newcommand{\be}{\begin{equation}}  
\newcommand{\ee}{\end{equation}}  
\newcommand{\bea}{\begin{eqnarray}}  
\newcommand{\eea}{\end{eqnarray}}  
\makeindex
\begin{document}

\vspace*{1.2cm}

\thispagestyle{empty}
\begin{center}
{\LARGE \bf Determination of Proton Parton Distribution}\\\vspace{2mm}
{\LARGE \bf  Functions using ATLAS Data}

\par\vspace*{7mm}\par

{

\bigskip

\large \bf Zhiqing Zhang, on behalf of the ATLAS Collaboration\footnote{Copyright 2021 CERN for the benefit of the ATLAS Collaboration. Reproduction of this article or parts of it is allowed as specified in the CC-BY-4.0 license.}}

\bigskip

{\large \bf  E-Mail: Zhiqing.Zhang@ijclab.in2p3.fr}

\bigskip

{Laboratoire de Physique des 2 infinis Ir\`ene Joliot-Curie $-$ IJCLab, Universit\'e Paris-Saclay, CNRS/IN2P3, 91405 Orsay, France}

\bigskip

{\it Presented at the Low-$x$ Workshop, Elba Island, Italy, September 27--October 1 2021}

\vspace*{15mm}

\end{center}
\vspace*{1mm}

\begin{abstract}

Using differential cross-section data of inclusive $W$ and $Z/\gamma^\ast$ boson production and of their production in association with jets, and differential cross-section data of top-quark pair production in the lepton + jets and dilepton channels in proton--proton collisions at the CERN Large Hadron Collider at center-of-mass energies of $\sqrt{s}=7$ and 8~TeV, together with deep inelastic scattering data from electron--proton collisions at the HERA collider, several determinations of parton distribution functions of the proton have been performed by the ATLAS experiment. The results are presented here. The emphasis is given to the determination of the strange sea quark distribution and its evolution over time.
\end{abstract}
 
 \section{Introduction}
Parton distribution functions (PDFs) of the proton cannot be predicted from first principle and they used to be determined in deep inelastic scattering (DIS) experiments using point-like beams of charged or neutral leptons to probe the proton or other nucleon targets, in earlier fixed target mode and later collision mode at HERA. Thanks to the factorization theorem, the PDFs determined from one process can be used as prediction for other processes. %The precision of the PDFs affects thus directly the accuracy of the theoretical predictions for cross sections and the interpretation of results obtained at the LHC.

At hadron colliders such as the Large Hadron Collider (LHC), the uncertainty of the PDFs is now the dominant uncertainty source for precision measurements, one example is the recent determination of the $W$ boson mass~\cite{1701.07240} by ATLAS~\cite{detector}. It also becomes a limiting factor for searches beyond the Standard Model. It is therefore important to improve our knowledge on PDFs by using relevant measurements at the LHC.

Since 2012, several analyses at next-to-next-to-leading order (NNLO) accuracy in quantum chromodynamics (QCD) have been performed by the ATLAS experiment using Drell-Yan or top-quark pair cross-section data as outlined in Table~\ref{tab:fits}. In the following sections, these QCD analyses and their corresponding results will be briefly described and shown.
\begin{table}
\centering
\caption{Overview of ATLAS QCD analyses and the used data sets including the neutral current (NC) and charged current (CC) data from HERA-I and HERA-I+II, and various ATLAS measurements of the Drell-Yan processes $W\to\ell\nu$ and $Z(\gamma^\ast)\to\ell\ell$ and of top-quark pair production at different center-of-mass energies.}
        \label{tab:fits}
        \resizebox{\columnwidth}{!}
        { 
          \begin{tabular}{c|cccc}
          \hline
          Data sets & epWZ12~\cite{1203.4051} & epWZ16~\cite{1612.03016} & epWZtop18~\cite{ATL-PHYS-PUB-2018-017} & epWZVjet20~\cite{2101.05095} \\\hline
          HERA-I NC, CC~\cite{0911.0884} & \checkmark &  &  &  \\
          HERA-I+II NC, CC~\cite{1506.06042} &  & \checkmark & \checkmark & \checkmark \\\hline
          ATLAS $W$, $Z$, 7 TeV 35 pb$^{-1}$~\cite{1109.5141}  & \checkmark & & & \\
          ATLAS $W$, $Z/\gamma^\ast$, 7 TeV 4.2 fb$^{-1}$~\cite{1612.03016} & & \checkmark & \checkmark & \checkmark \\\hline
          ATLAS $t\bar{t}$ $(\ell+$ jets, dilepton$)$ 8 TeV~\cite{1511.04716,1607.07281} & & & \checkmark & \\\hline
          ATLAS $W/Z+$ jets 8 TeV~\cite{1711.03296,1907.06728} & & & & \checkmark \\\hline
          \end{tabular}
        }
\end{table}
 
\section{Analysis ATLAS-epWZ12}

In 2012, a NNLO QCD analysis, ATLAS-epWZ12~\cite{1203.4051}, was performed by the ATLAS experiment to assess the strange quark distribution, using the first differential cross-section measurements of inclusive $W^\pm$ and $Z$ boson production based on a $pp$ collision data sample corresponding to an integrated luminosity of 35 pb$^{-1}$ recorded in 2010 at a center-of-mass energy of $\sqrt{s}=7$~TeV~\cite{1109.5141}, together with the combined $e^\pm p$ neutral current (NC) and charged current (CC) cross-section measurements of H1 and ZEUS from HERA-I~\cite{0911.0884}. 

The HERA measurements were included since they are the primary source for constraining PDFs with their large kinematic coverage for $Q^2$, the absolute four-momentum transfer squared, from near 1~GeV$^2$ to above $10^4$~GeV$^2$ and of Bjorken-$x$ from $\sim 0.6$ down to $10^{-4}$. However, they do have limitations. They cannot distinguish quark flavor between the down-type sea quarks, $\bar{d}$ and $\bar{s}$. Also the measurement precision at high $x$ and large $Q^2$ are statistically limited.

The ATLAS data, measured as functions of the $W$ decay lepton pseudorapidity, $\eta_\ell$, and of the $Z$ boson rapidity, $y_Z$, with a typical precision of $(2-3)\%$, access a kinematic range prescribed by the boson masses, $M_{W, Z}$, and the proton beam energy of $E_p=3.5$~TeV, corresponding to $Q^2\simeq M^2_{W,Z}$ and an $x$ range $0.001\lesssim x\lesssim 0.1$, with a mean $x=M_Z/2E_p=0.013$ for the $Z$ boson. The data provide new constraints on the strange quark distribution at high scale, $Q^2\sim M^2_{W,Z}$, which imply constraints at low $Q^2$ through perturbative  QCD evolution.

The QCD analysis used the HERAFitter framework~\cite{HERAFitter}. The light quark coefficient functions were calculated to NNLO as implemented in QCDNUM~\cite{1005.1481}. The contributions of heavy quarks are calculated in the general-mass variable-flavor-number scheme of Refs.~\cite{9709442,0601245}. The electroweak (EW) parameters and corrections relevant for the $W$ and $Z$ boson production processes were determined following the procedure described in Ref.~\cite{1109.5141} and cross-checked between the FEWZ~\cite{0312266} and the DYNNLO~\cite{0903.2120} programs. The fit package used APPLGRID code~\cite{0911.2985} interfaced to the MCFM program~\cite{1007.3492} for fast calculation of the differential $W$ and $Z$ boson cross sections at next-to-leading order (NLO) and a $K$-factor technique to correct from NLO to NNLO predictions. The data were compared to the theory using the $\chi^2$ function defined in Ref.~\cite{0904.0929}. 
%heavy quark scheme

The quark and gluon distributions at an initial scale of $Q^2_0=1.9$~GeV$^2$, chosen such that it is below the charm mass threshold $m_c$, were parameterized by generic forms:
\begin{eqnarray}
&& xq_i(x)=A_ix^{B_i}(1-x)^{C_i}P_i(x)\,,\nonumber\\
&& xg(x)=A_gx^{B_g}(1-x)^{C_g}P_g(x)-A^\prime_gx^{B^\prime_g}(1-x)^{C^\prime_g}\,,\nonumber
\end{eqnarray}
where $P_{i, g}$ denotes polynomials in powers of $x$. The fitted quark distributions were the valence quark distributions $(xu_v, xd_v)$ and the light quark sea distributions $(x\bar{u}, x\bar{d}, x\bar{s})$. 
The parameters $A_{u_v}$ and $A_{d_v}$ were fixed using the quark counting rule, $A_g$ using the momentum sum rule, and $C^\prime_g$ was set to 25 to suppress negative contributions at high $x$.  The normalization and slope parameters, $A$ and $B$, of $\bar{u}$ and $\bar{d}$ were set equal such that $x\bar{u} = x\bar{d}$ at $x\to 0$. For the strange quark distribution results presented here, it was parametrized with $P_{\bar{s}}=1$ and $B_{\bar{s}}=B_{\bar{d}}$. Terms were
added in the polynomial expansion $P_i(x)$ only if required by the data with a significant decrease in $\chi^2$ of the fit results, following the procedure described in Ref.~\cite{0911.0884}. This led to one additional term, $P_{u_v}=1+E_{u_v}x^2$, for the $u$ valence quark distribution, giving in total 15 free parameters for the fit.
The parton distributions at other $Q^2$ values are obtained using the evolution equations. 

\begin{figure}
\begin{center}
\includegraphics[width=0.7\columnwidth]{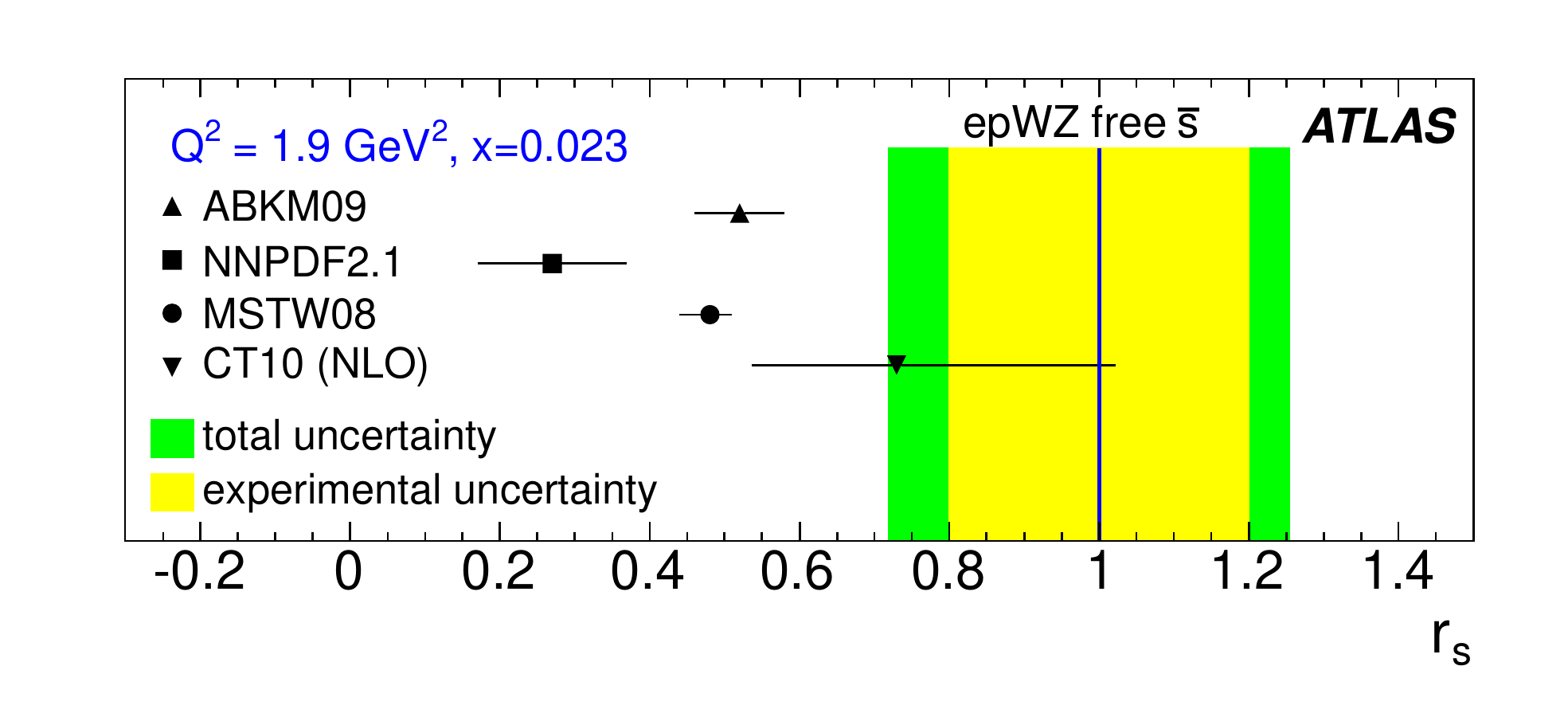}
\caption{Relative strange-to-down sea quark ratio for $Q^2_0=1.9$~GeV$^2$ and $x=0.023$ comparing the determination from ATLAS (shown in the vertical line with bands for experimental and total uncertainties) with predictions from different PDF sets (shown in closed symbols with horizontal error bars). The plot is taken from Ref.~\cite{1203.4051}.}
\label{fig:epwz12}
\end{center}
\end{figure}

The fit resulted in a good overall $\chi^2$ value per number of degree of freedom of 538.4/565 and determined the ratio of the strange quark distribution over the down quark one at $Q^2_0$ and $x=0.023$ (the $x$ value corresponds to 0.013 at $Q^2=M^2_Z$) to be:
\begin{equation}
r_s=\frac{s+\bar{s}}{2\bar{d}}=1.00\pm 0.20 (\textrm{exp})\pm 0.07 (\textrm{mod})^{+0.10}_{-0.15} (\textrm{par}) ^{+0.06}_{-0.07} (\alpha_s)\pm 0.08 (\textrm{th})\,,\nonumber
\end{equation}
where uncertainties are experimental, model, parametrization, $\alpha_s$ and theoretical. The determination, consistent with the prediction that the light quark sea at low $x$ is flavor symmetric, was compared in Figure~\ref{fig:epwz12} with predictions obtained from four global PDF determinations. The ABKM09~\cite{0908.2766} and MSTW08~\cite{0901.0002} determinations gave a value around 0.5 and the NNPDF2.1~\cite{1107.2652} determination was lower at about 0.25. On the other hand, the CT10 (NLO)~\cite{1007.2241} determination gave a large ratio consistent with the ATLAS determination.

\section{Analysis ATLAS-epWZ16}

The ATLAS-epWZ16 analysis~\cite{1612.03016} is very similar to ATLAS-epWZ12. The main differences are that the HERA-I $ep$ data were replaced with the final NC and CC DIS HERA-I+II data~\cite{1506.06042}. The ATLAS Drell-Yan measurements used~\cite{1612.03016} were based on the full 7~TeV  data sample of 4.6 fb$^{-1}$ with improved precision. In addition, the measurement of the $Z/\gamma^\ast$ process was extended in three mass ranges of $46<m_{\ell\ell}<66$~GeV, $66<m_{\ell\ell}<116$~GeV and $116<m_{\ell\ell}<150$~GeV for the central rapidity region of $|y_{\ell\ell}|<2.5$ and in two mass ranges $66<m_{\ell\ell}<116$~GeV and $116<m_{\ell\ell}<150$~GeV in the forward rapidity region up to 3.6. The NNLO QCD fit with the same PDF parameterization as the ATLAS-epWZ12 fit was performed using the xFitter program~\cite{xFitter}.

\begin{figure}
\begin{center}
\includegraphics[width=0.55\columnwidth]{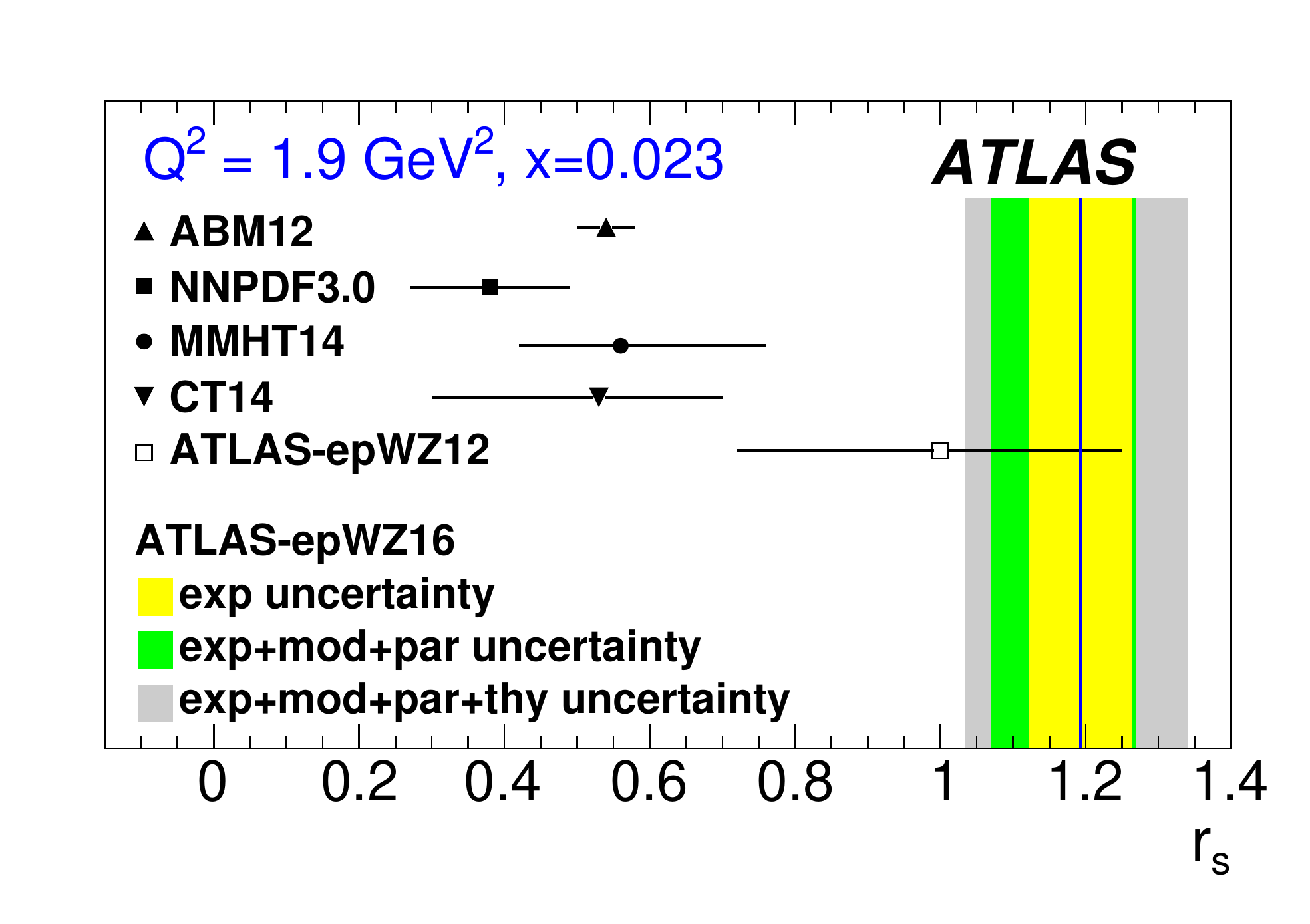}
\includegraphics[width=0.4\columnwidth]{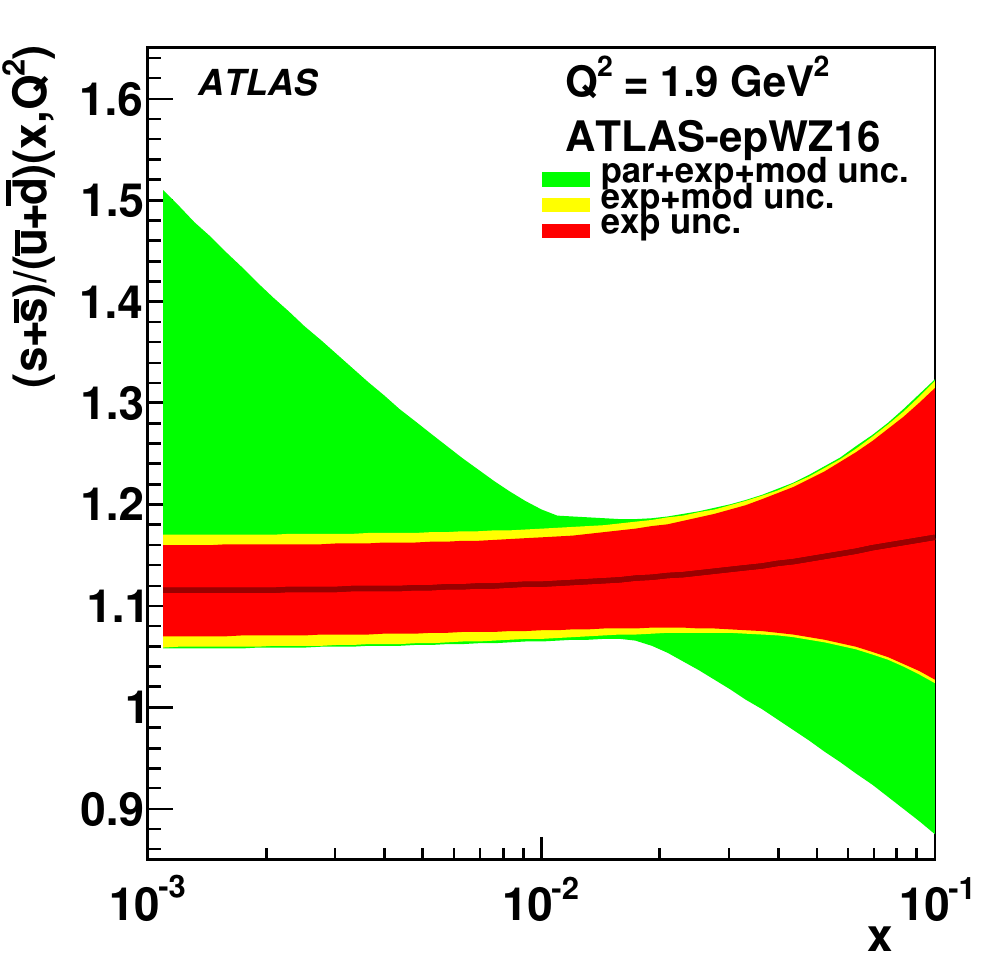}
\caption{Left: Relative strange-to-down sea quark ratio $r_s$ for $Q^2_0=1.9$~GeV$^2$ and $x=0.023$ comparing the determination from ATLAS (shown in the vertical line with bands for experimental data, QCD fit and theoretical uncertainties) with predictions from different NNLO PDF sets (shown in closed symbols with horizontal error bars) as well as previous ATLAS results (shown in open square). Right: relative strange quark ratio over the sum of up and down sea quarks as a function of Bjorken-$x$. The plots are taken from Ref.~\cite{1612.03016}.}
\label{fig:epwz16}
\end{center}
\end{figure}

\sloppy
The strange quark ratio $r_s$ was determined with improved precision at $Q^2_0$ and $x=0.023$:
\begin{equation}
r_s=1.19\pm 0.07 (\textrm{exp})^{+0.13}_{-0.14} (\textrm{mod}+\textrm{par}+\textrm{th})\,.\nonumber
\end{equation} 
The comparison with predictions from other global PDF sets ABM12~\cite{1310.3059}, NNPDF3.0~\cite{1410.8849}, MMHT14~\cite{1412.3989}, CT14~\cite{1506.07443}, as well as from ATLAS-epWZ12 is shown in Figure~\ref{fig:epwz16} (left). In addition, the ratio of strange quarks over the sum of up and down sea quarks $R_s=(s+\bar{s})/(\bar{u}+\bar{d})$ as a function of $x$ has also been obtained (Figure~\ref{fig:epwz16} (right)), though the uncertainty in particular the parametrization uncertainty was large.

\section{Analysis ATLAS-epWZtop18}

Another NNLO QCD analysis, ATLAS-epWZtop18~\cite{ATL-PHYS-PUB-2018-017}, included, in addition to those data used in the ATLAS-epWZ16 fit, top-quark pair production data measured in the lepton + jets~\cite{1511.04716} and dilepton~\cite{1607.07281} channels at 8~TeV corresponding to an integrated luminosity of 20.2~fb$^{-1}$. The top-quark pair production data are complementary to the other data sets in their PDF constraining power since they are expected to be sensitive to the high-$x$ gluon distribution~\cite{1611.08609}. The corresponding NNLO prediction~\cite{1704.08551} were supplied in the form of FastNLO grids~\cite{1208.3641} for the data in the lepton + jets channel. For the dilepton channel, APPLGRID was interfaced to MCFM to produce NLO grids and a $K$-factor technique was used to correct from NLO to NNLO predictions. 

% the form of the central parameterisation has one more free parameter (Du¯ ) as compared to the ATLASepWZ16 analysis [3] and a total of 16 free parameters.

The setup of the fit differs from that of ATLAS-epWZ16 in three aspects, apart from the addition of top-quark pair production data. Firstly, the low mass off-peak $Z/\gamma^\ast$ boson data was not used because it became clear that the lower $x$ region probed by the data was subject to further QCD~\cite{1710.05935} corrections which were not readily calculable, this had negligible impact on the fits. Secondly, the minimum $Q^2$ selection on the HERA data was raised from 7.5~GeV$^2$ to 10~GeV$^2$, this larger cut was already considered as one of the model variations of the ATLS-epWZ16 fit and was used as standard in the new analysis to avoid the region of the HERA data which may be subject to $\ln(1/x)$-resummation effects~\cite{1710.05935} that were not accounted for in the prediction. Thirdly, the ATLAS-epWZtop18 fit had one additional term, $P_{\bar{d}}=1+D_{\bar{d}}x$, for the down sea quark distribution than those of ATLAS-epWZ16 where it was considered as a parameterization variation.

\begin{figure}
\begin{center}
\includegraphics[width=0.6\columnwidth]{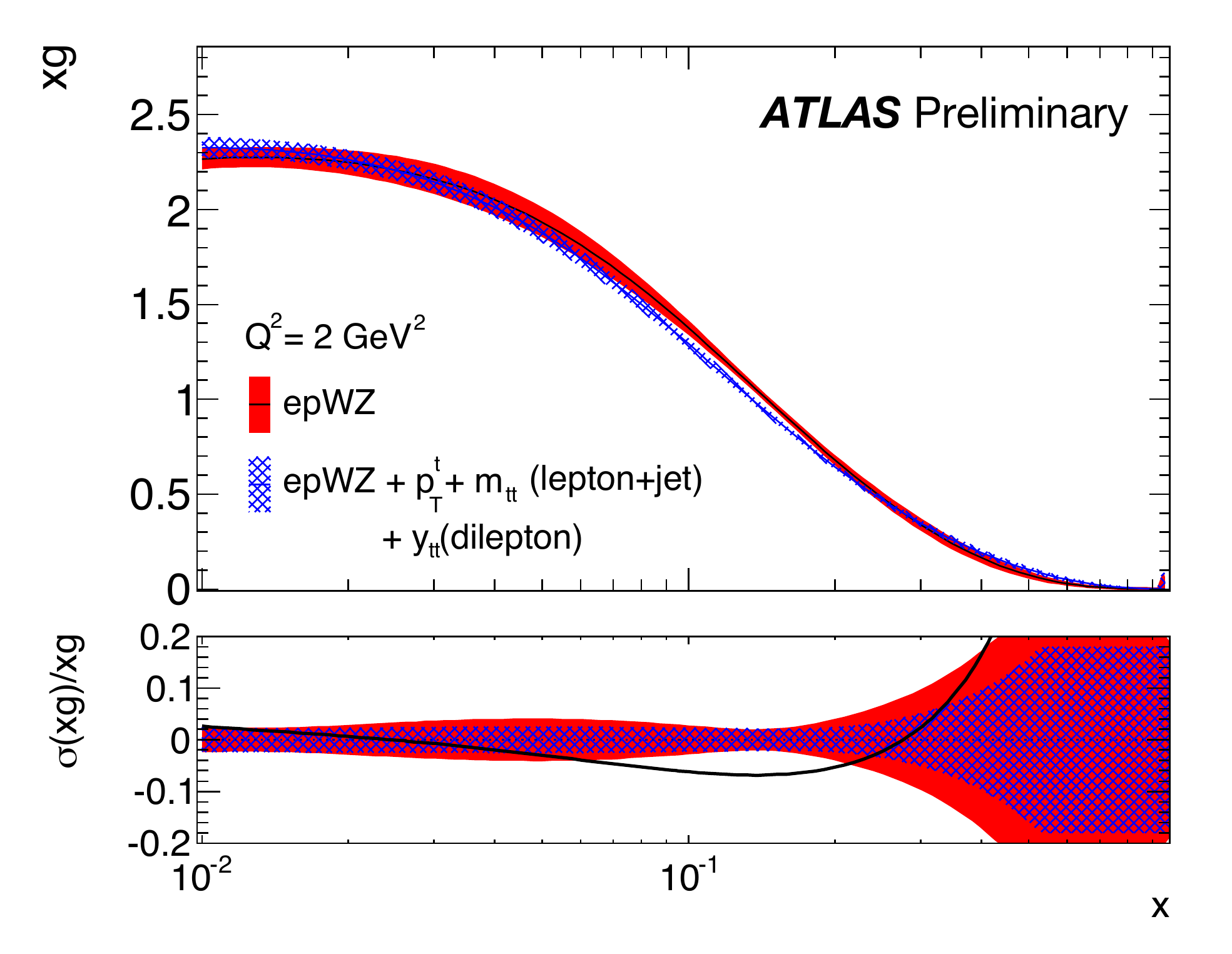}
\caption{Gluon distribution as a function of Bjorken-$x$ from fitting HERA data and ATLAS $W$ and $Z/\gamma^\ast$ boson data plus the $t\bar{t}$ lepton + jet  $p^t_\textrm{T}$ and $m_{t\bar{t}}$ data and the $t\bar{t}$ dilepton $y_{t\bar{t}}$ data compared to the fit to HERA and ATLAS $W$ and $Z/\gamma^\ast$ boson data alone. The plot is taken from Ref.~\cite{ATL-PHYS-PUB-2018-017}.}
\label{fig:top18}
\end{center}
\end{figure}

The main result of the fit is presented in Figure~\ref{fig:top18}, showing the impact of the top-quark pair production data on the gluon distribution; reducing its uncertainty in particular at high $x$ and making it softer and harder at medium $x$ and high $x$, respectively.

\section{Analysis ATLAS-epWZVjet20}

The analysis ATLAS-epWZVjet20~\cite{2101.05095} included new ATLAS data on the production of a vector ($W$ or $Z$) boson in association with at least on jet~\cite{1711.03296,1907.06728}. The measurements were based on the same data sample as for the top-quark pair production at 8~TeV. They provided a novel source of input to PDF determination that is sensitive to partons at higher $x$ and $Q^2$ than can be accessed by $W$ and $Z$ boson data alone. These data are thus complementary to the inclusive $W/Z$ boson measurements. %The tree-level production modes of a vector boson in association with jets ($V +$ jets) have either quark–antiquark initial states with gluon radiation, or quark–gluon initial states. The process is therefore already sensitive to the gluon density of the proton at leading order in QCD, while providing constraints on the quark distributions in a similar way to inclusive production of a vector boson.

Predictions for $W +$ jets and $Z +$ jets production~\cite{1511.08692} were obtained similarly to the $W$ and $Z$ predictions to NLO in QCD and leading order in EW couplings by using the APPLGRID code interfaced to the MCFM program. NNLO (NLO) corrections in QCD (EW) were implemented as $K$-factors. 

\begin{figure}
\begin{center}
\includegraphics[width=0.49\columnwidth]{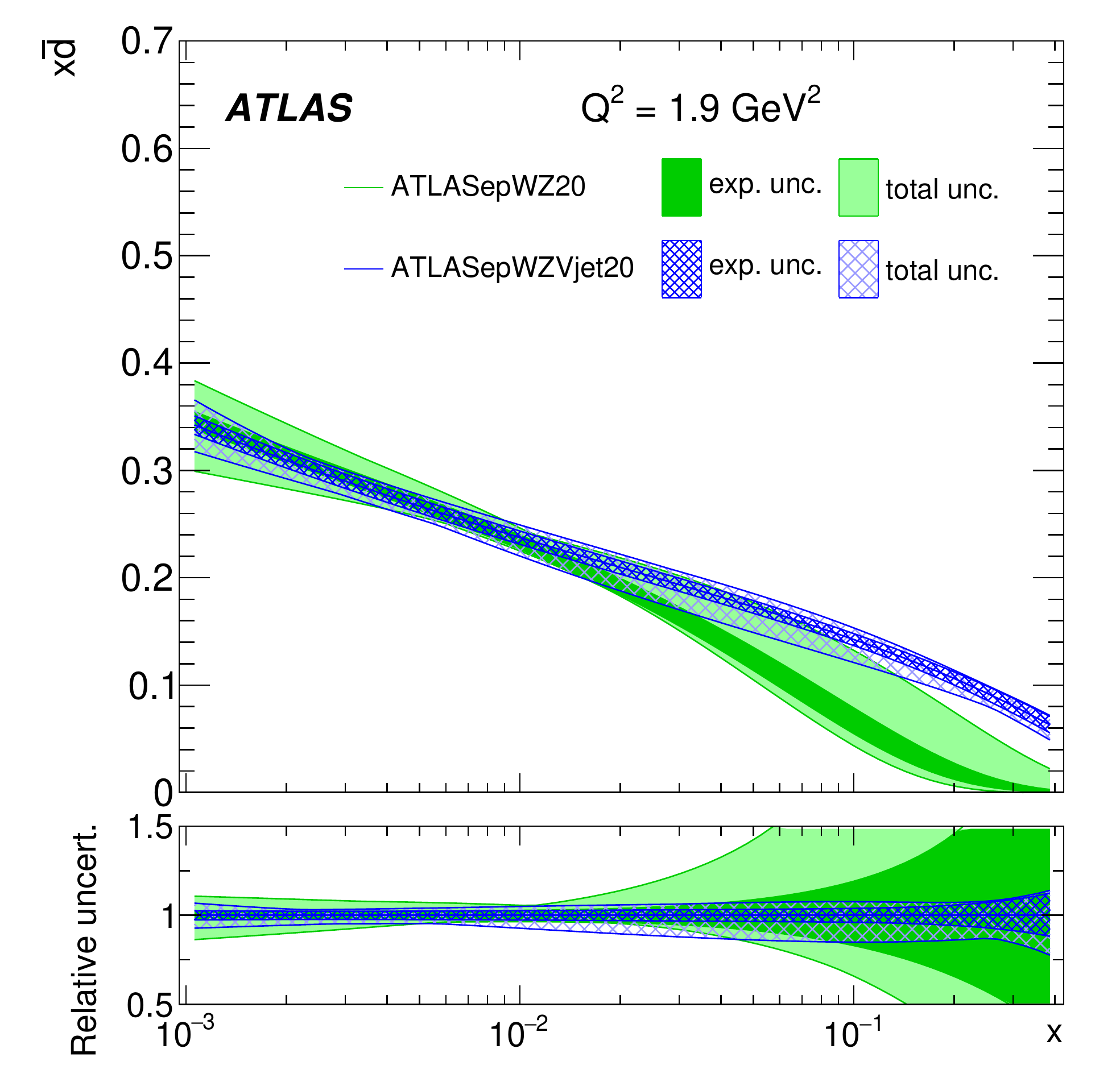}
\includegraphics[width=0.49\columnwidth]{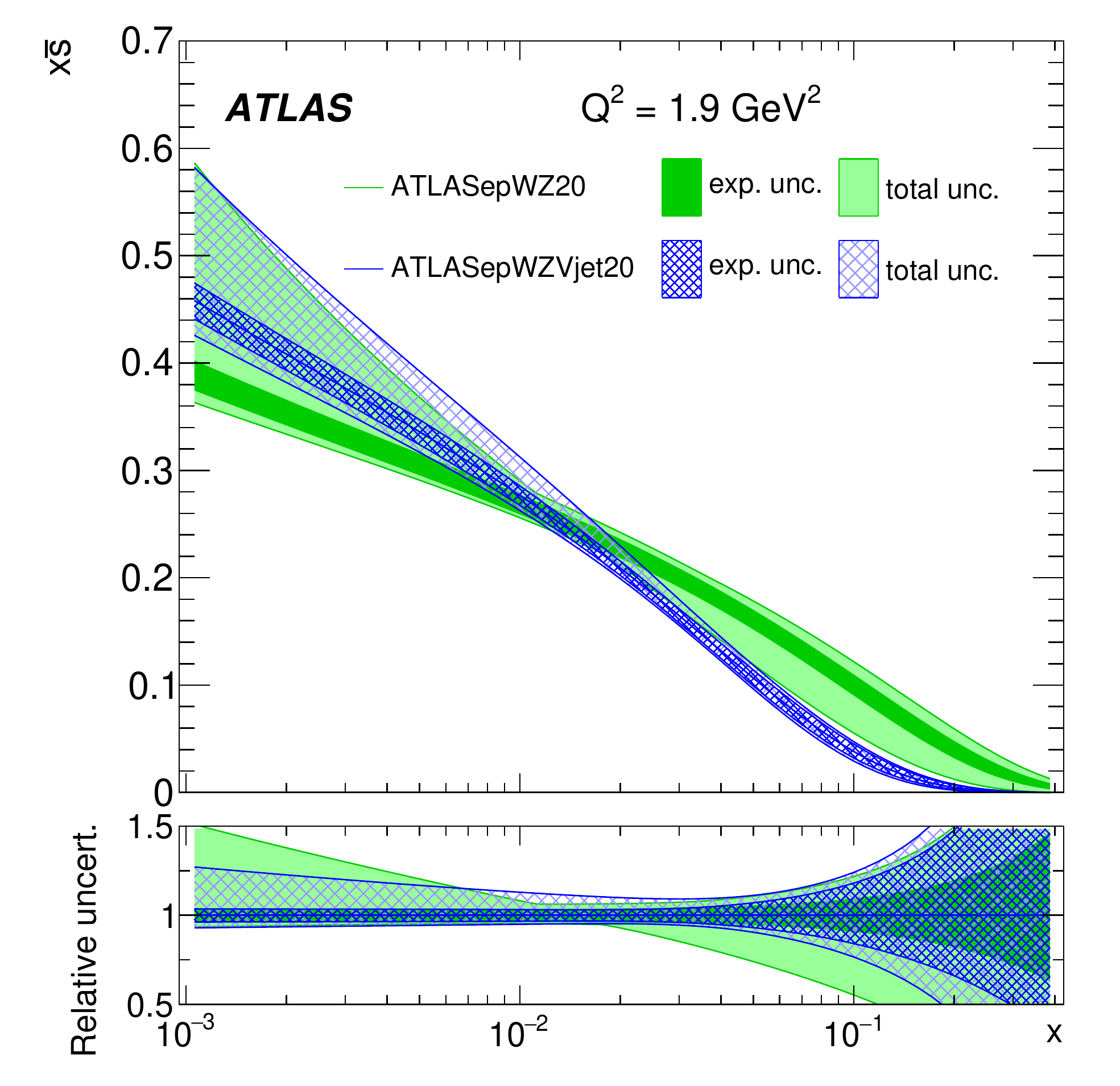}
\includegraphics[width=0.49\columnwidth]{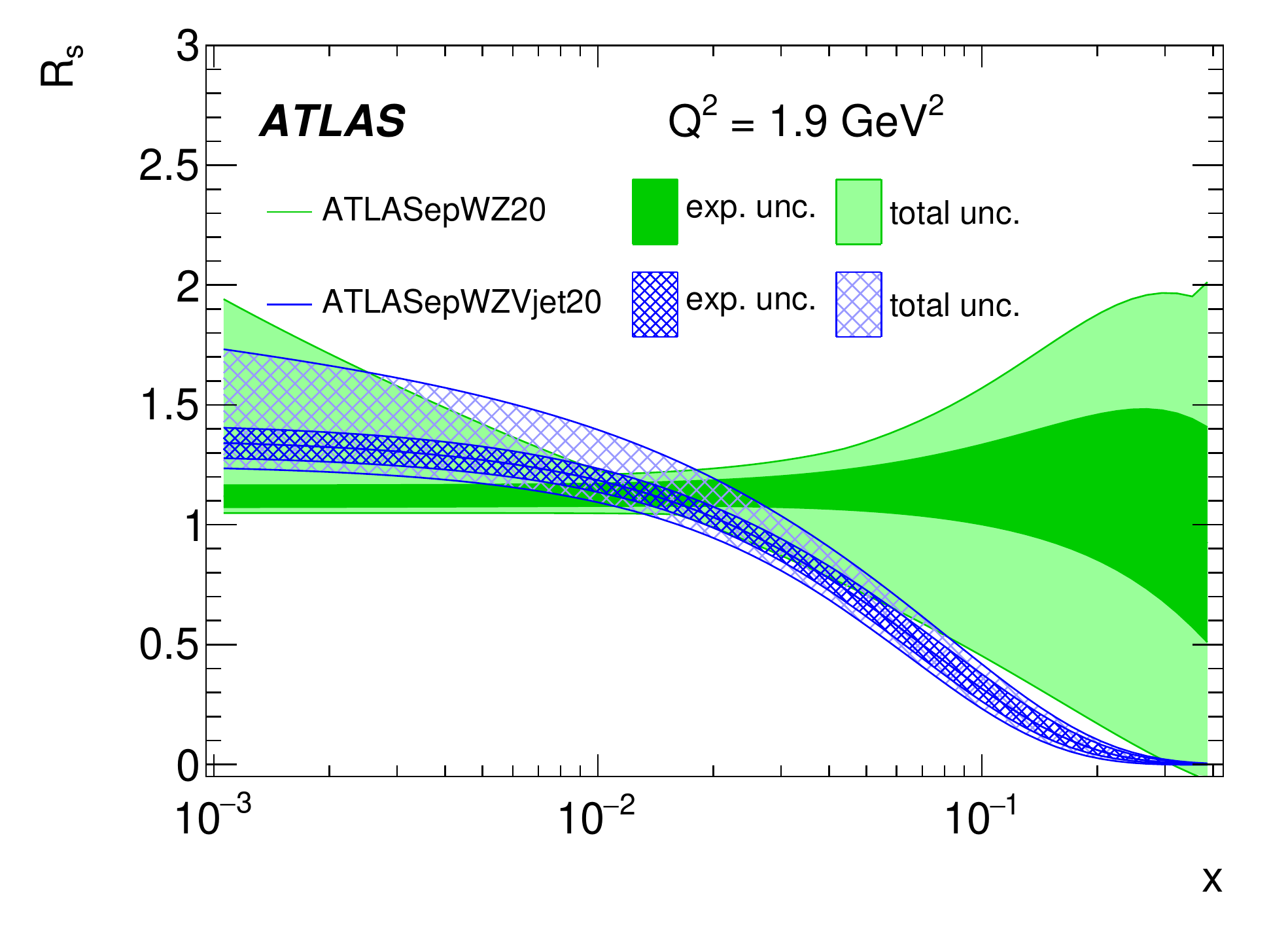}
\includegraphics[width=0.49\columnwidth]{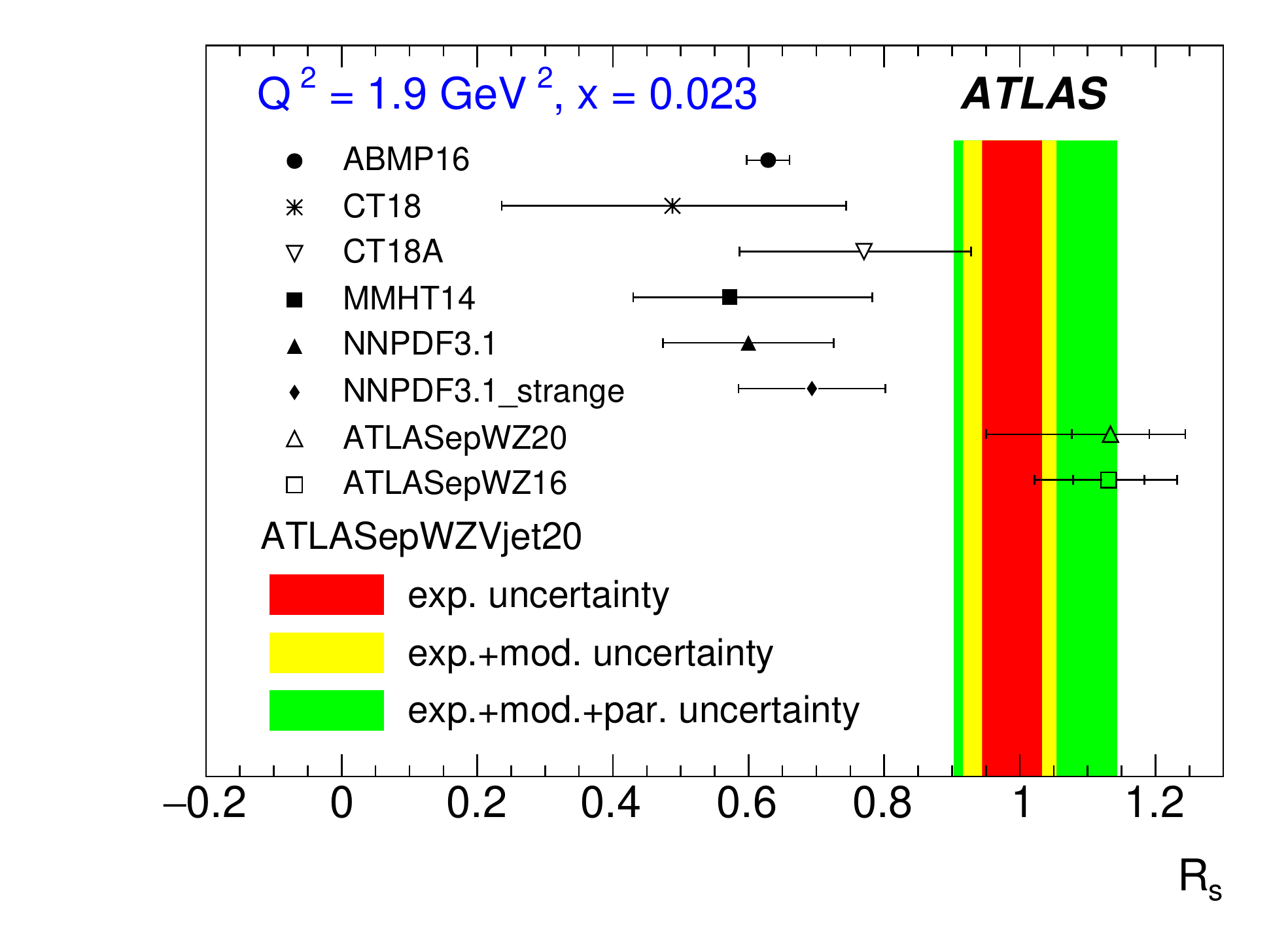}
\caption{Top-left: Down sea quark distribution as a function of Bjorken-$x$ when including (blue) or without including (green) the $V+$ jets data in the fits. Top-right: similar plot for strange sea quark distribution. Bottom-left: similar plot for relative strange quark ratio over the sum of up and down sea quarks $R_s$ as a function of Bjorken-$x$. Bottom-right: $R_s$ determination in comparison with other predictions for $Q^2_0=1.9$~GeV$^2$ and $x=0.023$. The plots are taken from Ref.~\cite{2101.05095}.}
\label{fig:ds_vjet20}
\end{center}
\end{figure}

The setup of the ATLAS-epWZVjet20 fit differs from that of ATLAS-epWZ16 in that there was one more term, $P_g=1+D_gx$, for the gluon distribution and a tighter selection of $Q^2>10$~GeV$^2$ instead of 7.5~GeV$^2$ on the DIS data. The resulting down and strange sea quark distributions of the fit and $R_s$ at the starting scale $Q^2_0$ are shown in Figure~\ref{fig:ds_vjet20} in comparison with the results of the ATLAS-epWZ20 fit when the $V+$ jets data were not included. The inclusion of the $V+$ jets data made the $x\bar{d}$ distribution notably higher in the range of $x\gtrsim 0.02$ and the $x\bar{s}$ distribution lower in the same range, with significantly smaller experimental and parameterization uncertainties at high $x$. The difference between the two $R_s$ determinations at high $x$ is essentially covered by the large uncertainty dominated by the parameterization uncertainty in the ATLAS-epWZ20 fit. The new $R_s$ determination at $x=0.023$ was again compared with the predictions from ABMP16~\cite{1701.05838}, CT18 and CT18A~\cite{1912.10053}, MMHT14~\cite{1412.3989}, NNPDF3.1 and NNPDF3.1\_strange~\cite{1706.00428} and the ATLAS determinations ATLAS-epWZ20 and ATLAS-epWZ16, the latter two being different mainly in their PDF parameterizations. Better agreement was observed with the CT18A PDF set, which included both the data used in the CT18 fit and the ATLAS 7 TeV data, although tension remains with the NNPDF3.1\_strange PDF set, which also used this data.

\begin{figure}
\begin{center}
\includegraphics[width=0.5\columnwidth]{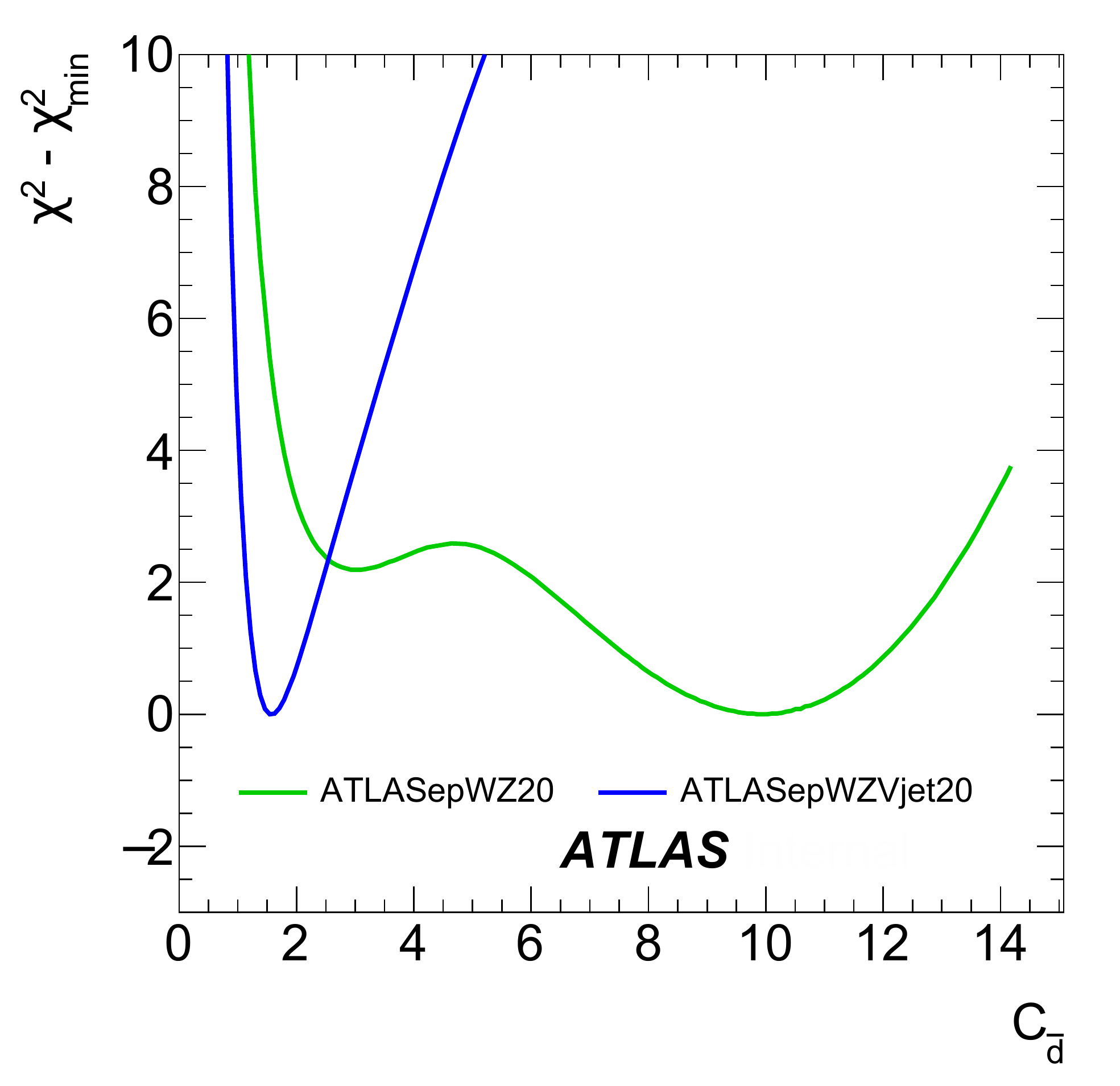}
\caption{Variation of $\chi^2$ subtracting the minimum value when including (blue) or without including (green) the $V+$ jets data in the fits as a function of parameter $C_{\bar{d}}$ . The plot is taken from Ref.~\cite{2101.05095}.}
\label{fig:dm_vjet20}
\end{center}
\end{figure}

The large change at high $x$ was investigated by performing a $\chi^2$ scan of the $C_{\bar{d}}$ parameter which controls the high-$x$ behavior of the $x\bar{d}$ distribution. The $\chi^2$ variation subtracting the minimum value is presented in Figure~\ref{fig:dm_vjet20}. The other data sets showed double minima in $\chi^2$ with a preferred solution for a $C_{\bar{d}}$ value around 10 corresponding to a soft $x\bar{d}$ distribution at high $x$. The inclusion of the $V+$ jet data allowed to resolve the ambiguity and resulted in a $C_{\bar{d}}$ value around 1.6 and thus a harder $x\bar{d}$ distribution at high $x$. Since the sum of $x\bar{d}+x\bar{s}$ is well constrained by e.g.\ the DIS data, the harder $x\bar{d}$ distribution gives thus the softer $x\bar{s}$ distribution and smaller $R_s$ at high $x$.

\FloatBarrier

\section{Summary and prospect}

The results of several NNLO QCD analyses performed using the Drell-Yan production and top-quark pair production cross sections measured by ATLAS with $pp$ collision data at the LHC taken at center-of-mass energies of 7 and 8~TeV, together with the neutral and charged current deep inelastic scattering cross sections from $ep$ collisions at HERA, have been presented. The analyses show the constraining power of the ATLAS measurements on parton distribution functions of the proton, complementary to the HERA data. In particular, the strange quark distribution at low $x$ is found unsuppressed contrary to what was assumed and obtained in the other PDF analyses. The shape of the strange quark ratio over the other light sea quarks at high $x$ has been better understood thanks to the measurement of vector-boson production in association with at least one jet.

Another valuable constraint for the strange quark distribution at the LHC concerns the associated production of $W$ bosons and charm quarks since it probes the strange quark content of the proton directly through the leading order processes $g + \bar{s} \to W^+ +\bar{c}$ and $g + s \to W^- +c$. Earlier analyses, not presented here, performed NLO fits using the measurements of ATLAS at 7~TeV~\cite{1402.6263} and of CMS at 7~TeV~\cite{1310.1138} and at 13~TeV~\cite{1811.10021}. With the recent available NNLO predictions~\cite{2011.01011}, there is a good prospect in including the $W+c$ data in the future NNLO analyses, though the interpretation of the data is sensitive to the modelling of $c$-quark fragmentation.

%allowing the strange content of the sea to be fitted, rather than assumed to be a fixed ratio of the light sea as is required when fitting HERA inclusive data alone.

%\section{Results}
%We describe here the details our results.

%The document should not exceed 10 pages excluding references (please contact the organizers if you need more pages). The proceedings will be published online for free by the University of Kansas (this will be an "official" publication) and printed copies will be made available for a reasonable cost. The deadline to send the \textbf{.tar} file of the contribution (\textbf{including the .pdf} which means a successful compilation) is December 15 2021 to \textbf{christophe.royon@ku.edu, lalcerro@ku.edu, gkrintir@ku.edu}. It is also possible to put the article on arXiv provided you mention on the relevant field 

%Comments: Presented at the Low-$x$ Workshop, Elba Island, Italy, September 27--October 1 2021.

%\section*{Acknowledgements}

%The author thanks somebody.

\end{document}